# Dynamics of DNA-Templated Ultrafine Silver Nanowires Formation

Qifei Ma,[a] Mauro Chinappi,[b] Ali Douaki,[c] Yanqiu Zou,[d] Huaizhou Jin,[e]* Emiliano Descrovi,[f] Roman Krahne,[c] Remo Proietti Zaccaria,[c] Dan Cojoc,[g] Karol Kolataj,[h] Guillermo Acuna,[h] Shangzhong Jin[a]* and Denis Garoli*[a,c,i]

Recent research on silver nanowires prepared on DNA templates has focused on two fundamental applications: nano-scale circuits and sensors. Despite its broad potential, the formation kinetics of DNA-templated silver nanowires remains unclear. Here, we present an experimental demonstration of the formation of silver nanowires with a diameter of 2.2+0.4 nm at the single-molecule level through chemical reduction. We conducted equilibrium and perturbation kinetic experiments to measure force spectroscopy during the formation of $Ag^+$-DNA complexes and Ag-DNA complexes, using optical tweezers combined with microfluidics. The addition of $AgNO_3$ resulted in an increase in force of 5.5-7.5 pN within 2 minutes, indicating that $Ag^+$ compacts the DNA structure. In contrast, the addition of hydroquinone caused the force to decrease by 4-5 pN. Morphological characterization confirmed the presence of a dense structure formed by silver atoms bridging the DNA strands, and revealed conformational differences before and after metallization. We compare our experimental data with Brownian dynamics simulations using a coarse-grained double-stranded DNA (dsDNA) model that provides insights on the dependency of the force on the persistence length.

Keywords: force spectroscopy, single-molecule, DNA, optical tweezers, metallization

## Introduction

Silver nanowires have attracted attention due to their high electrical conductivity[1,2] and interesting optical properties[3–7] that promise great potential in nanoscale circuits[8–11], sensors[12–20]. Braun et al. [21] first proposed a conducting silver nanowire templated by DNA, which overcomes poor electrical characteristics of DNA molecules and might allow the application of DNA to construct functional circuits.  The morphological, electrical and optical characteristics and stability of silver nanowires are significantly influenced by the effectiveness of their protection groups (i.e. surface functionalization) that shields them from external environmental factors. For instance, the luminescence of silver can either increase or decrease due to external conditions, highlighting the potential for utilizing silver nanowires as sensitive optical detectors for biothiols[13,22], gas[14,17], DNA[23–25] and other substances. Additionally, luminescent silver nanowires are efficient emitters with high quantum yields, high molecular extinction coefficients, nanosecond lifetimes and excellent photostability, which can be exploited for visualization of genomic DNA sequences[12]. All of these applications rely on the interactions between DNA, silver ions ($Ag^+$) and silver atoms (Ag), and therefore insights into the mechanisms of DNA metallization and control the formation of the silver coating and the morphology of the resulting nanowires are still of important interest.

To date, relatively few experimental studies[26–32] have focused on the interaction between $Ag^+$ and DNA. Current understanding is that $Ag^+$ ions favor coordination with DNA bases over electrostatic binding to DNA phosphates,[26] and this coordination with DNA bases can be classified based on the Ag/DNA nucleotide ratio in three types of  binding modes.[27] Below an Ag/nucleotide ratio of 0.2, the type I complex is predominantly formed, with the guanine N7 atom as the key binding site. At ratios between 0.2 and 0.5, type II complexes are produced, primarily binding to A-T or G-C base pairs. At ratios greater than 0.5, type III complexes develop once the main binding sites in type I and type II complexes are fully saturated. Petty et. al[3] explored small (<10 atoms) Ag nanoclusters templated by DNA for the time-dependent and size-specific formation of nanoclusters by spectrum and mass spectrometry. They revealed that silver nanoclusters induce different structural changes in DNA as compared to silver ions. Sufficiently small silver nanoclusters exhibit very strong absorption and emission, making them nearly ideal fluorophores for single molecule spectroscopy, suggesting potential utility as biological labels.

a) College of Optical and Electronic Technology, China Jiliang University, Hangzhou 310018, China;
b) Department of Industrial Engineering, University of Rome Tor Vergata, Italy
c) Istituto Italiano di Tecnologia, Via Morego 30, 16163 Genova, Italy.
d) College of Optical Science and Engineering, Zhejiang University, Hangzhou 310027, China.
e) Key Laboratory of Quantum Precision Measurement, College of Physics, Zhejiang University of Technology, Hangzhou, China.
f) Dipartimento di Scienza Applicata e Tecnologia, Politecnico di Torino, Corso Duca degli Abruzzi 24, 10129 Torino, Italy
g) CNR-Istituto Officina dei Materiali (CNR-IOM),SS 14 km 163.5, Area Science Park Basovizza,34149, Trieste, Italy
h) Department of Physics, University of Fribourg, Fribourg CH 1700, Switzerland.
i) Dipartimento di Scienze e metodi dell'ingegneria, Università degli Studi di Modena e Reggio Emilia, 42122 Reggio Emilia, Italy.

* Corresponding authors
‡ E-mail Address: 18367198456@163.com;  jinsz@cjlu.edu.cn; denis.garoli@unimore.it.

Zinchenko et.al. [33] and Yang et.al. [34] focused on the conformational changes caused by the formation of silver nanoparticles based on DNA as a template and discussed the conformational behavior of a long single-chain double-stranded DNA in solutions of free silver ions and silver nanoparticles. Silver nanowires with few nm diameter are interesting for nano-scale circuits that go beyond the resolution limit of standard top-down fabrication processes.

In this work, we used a two-step chemical reduction method[21] to prepare the DNA-templated Ag nanowires. First, $Ag^+$ was added to the DNA to form $Ag^+$-DNA complexes. Then, hydroquinone was added to reduce $Ag^+$ to Ag atoms, resulting in the formation of the nanowires. By adjusting the concentration of DNA, $Ag^+$ and hydroquinone, we demonstrate the formation of a 2.2 nm silver nanowire at the single-molecule level and investigate the growth dynamics by real-time force spectroscopy measurements. Our study reveals that $Ag^+$ ions have a sensitive impact on DNA conformation, while Ag nanoclusters showed minimal interaction with DNA. By performing atomic force microscopy (AFM) imaging, we track the changes of the DNA conformation and the coverage of metal ions and metal atoms on DNA during the formation of the silver wires. We support our experimental findings by analyzing the dependency of the force on the persistence length with Brownian simulations of a coarse-grained dsDNA model.

## Results and Discussion

### *Stretching single DNA molecules and AFM characterization*

To study the effect of silver ions and silver atoms on the force change of an individual DNA molecule, a single λ-DNA molecule was suspended in the PBS buffer by optical tweezers (OTs) at room temperature.

The distance between the two beads and the stretching force are related. First, the distance (*d*) between the two beads is adjusted so that the initial force is about 2 pN (Fig. 1) (the importance of the initial force is discussed in Supporting Information – note#1). During metallization, the distance between the two beads is held constant, and the change in force is measured throughout the process. The experimental procedure involved the sequential addition of reagents. Initially, the introduction of a 0.05 mM $AgNO_3$ solution induced the formation of a DNA-$Ag^+$ complex, resulting in a force increase of 5.5–7.5 pN (Fig. 1b). This indicates that the addition of $Ag^+$ ions causes DNA compaction (*d*>*d₁*, Fig. 1a), and to maintain a constant distance between the two beads, a greater force is required to stretch the DNA. Subsequently, the addition of a 0.002 mM hydroquinone solution facilitated the formation of the DNA-Ag complex, leading to a force decrease of 4–5 pN (Fig. 1c-e). The production of silver atoms weakens the compaction of the DNA (*d*>*d₂*>*d₁,* Fig. 1a), thereby reducing the pulling force compared to the force observed after the addition of $Ag^+$ ions. Fig. 1b shows force stability after the addition of $Ag^+$ for 60 minutes. Within the first 2 minutes, $Ag^+$ fully bind to DNA, and over a period of up to 60 minutes, the structure of the $Ag^+$-DNA complex remains stable, demonstrating that the binding is strong and durable.

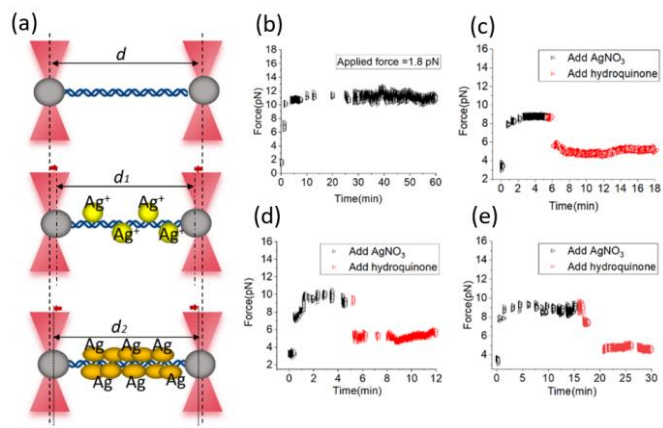

**Figure 1**. (a) The process of the force measurement during DNA metallization by OTs. (b)Force stability after addition of $Ag^+$ for 60min. (c)(d)(e) Force-time curves of λ-DNA in the presence of $Ag^+$ (black)and hydroquinone(red) from three tethers, respectively.

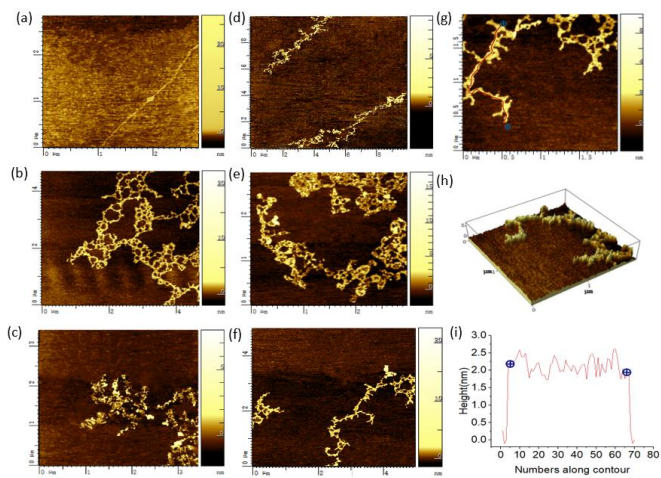

**Figure 2**. AFM images showing the morphology of (a ) single λ-DNA; (b-d) λ-DNA reacted with 0.05 mM $AgNO_3$ for 10, 20, 30 minutes, respectively ;(e-g) silver coated DNA nanowires as prepared in (b-d) with an additional incubation of $Ag^+$-DNA complex in 0.002 mM hydroquinone for 10 minutes; (h-i) Morphology images and height profile of λ-DNA and 0.05 mM $AgNO_3$ after 20 minutes incubation time + incubation time of $Ag^+$-DNA complex in 0.002 mM hydroquinone.

In order to have a direct demonstration of the DNA metallization, and to verify the growth rate of Ag ions, we performed AFM characterizations using similar conditions to the ones used in the tweezer experiments (Fig. 2). To note, in the OT experiment the reaction time is shorter than in the AFM experiment, because the microfluidic volume and the ratio of $Ag^+$ ions to DNA is much higher in the OT setup. As demonstrated in fig. 1c-e, the force-time curves of λ-DNA in the presence of $Ag^+$ (black) and hydroquinone(red) show that within the first 2 minutes, $Ag^+$ ions were reduced to Ag, forming an Ag nanowire, which was further evidenced by AFM morphology characterizations (Fig. 2). The AFM also enabled to demonstrate (Fig. 2i and Supporting Information Fig. S2) the good control regarding the final height of metallized DNA that for this structure was 2.2+0.4 nm.

The distinct coordination chemistries of $Ag^+$ ions and Ag atoms with DNA structural elements are at the origin of the opposite conformational effects that we observe. Previous studies have demonstrated that $Ag^+$ interacts strongly with nitrogen atoms in DNA bases, especially in GC base pairs[35–40]. This specific binding is considered to induce base pair rearrangements, which could be the main reason for the significant reduction in the persistence length of DNA. In contrast to $Ag^+$, the relatively weak interactions[40] between Ag atoms and oxygen atoms in DNA permit the DNA to regain its native conformation. This observation is consistent with the circular dichroism spectroscopic studies conducted by Petty et al.[3], which demonstrated that silver nanoclusters induce different structural changes in DNA compared to $Ag^+$. During the reduction process of $Ag^+$, the mechanical force did not fully return to its initial value, suggesting that the formation of small Ag clusters within the DNA structure partially hindered the complete reduction of $Ag^+$. Fig. 2a shows the AFM image obtained scanning a single λ-DNA. Because of the long chain of the λ-DNA (full length of about 16 μm), only part of the chain is shown with a measured height of about 1.5 nm, in agreement with previous reports[41,42]. Extended incubation of $Ag^+$ with DNA resulted in the localized formation of large metallic aggregates, which we attribute to the persistent coil state of DNA maintained by $Ag^+$ incorporation. This leads to an increased local charge density in specific DNA regions, thereby facilitating the accumulation and progressive growth of larger $Ag^+$ clusters through enhanced electrostatic interactions, consistent with reports by Zinchenko et al.[33] However, subsequent addition of hydroquinone destabilized the coiled conformation of the $Ag^+$-DNA complex, facilitating the uniform distribution of Ag atoms along the entire DNA strand. Since in the OT experiments the DNA molecule maintained a linear conformation, it prevented the formation of localized large-scale metallic aggregates.

Yang et al.[34] investigated the reductant dependent DNA-templated silver nanoparticle formation using magnetic tweezers. Their findings revealed that $Ag^+$ ions induce a partially condensed coil state in DNA strands, which is consistent with our OT results. However, their study also demonstrated that the formation of fully compact silver nanoparticles can be controlled by varying the types of reducers and the $Ag^+$/reducer ratio. In our work we focused on a complementary process, which allows to fabricate of silver nanowires from $Ag^+$ functionalized DNA by the addition of 0.05 mM $Ag^+$ and 0.002 mM hydroquinone.

As an additional investigation of the metallization process, we performed a set of Raman spectroscopies on the prepared DNA+Ag using the same preparation conditions used in the AFM experiments. The results are reported in Supporting Information note#3. As expected, the addition of Ag ions and the growth of small metallic particles/clusters on the DNA produces localized electromagnetic fields and enhances the Raman signals.

*DNA model*

To interpret the observed increase in the force due to the $Ag^+$ ions, we rely on established knowledge of polymer physics. Several coarse-grained models have been developed for DNA, and some are better suited for fitting experimental data on stretched DNA [43]. However, for our purposes, the simple ideal chain (or freely-jointed chain) model is appropriate to explain the trends in our data. In this model, the polymer consists of N rigid elements of fixed length b, the Kuhn length, so that the total length of the polymer (contour length) is L = Nb. Defining the end-to-end vector R as the vector pointing from the first to the last monomer, the free energy of an ideal chain is given by:

$$A(N, \mathbf{R}) = \frac{3k_BT}{2Nb^2}R^2 \quad (1)$$

with $k_B$ the Boltzmann constant and $T$ the temperature[44]. The free-energy of the chain increases quadratically with end-to-end distance $R = |\mathbf{R}|$. Consequently, the chain can be considered a harmonic entropic spring, where the spring elongation corresponds to the end-to-end distance R, and the force is given by:

$$\mathbf{F} = -k_e\mathbf{R} \; ; \; k_e = \frac{3k_BT}{Nb^2} \quad (2)$$

In our set-up, the DNA is bound to two beads at a distance $L_0$ whose positions are controlled by Ots (see Fig. 3). Thus, the end-to-end distance is $L_0$, and for the entropic spring model, the DNA exerts a force on the two beads proportional to their distance $L_0$. The interaction between dsDNA and Ag+ is expected to compact the DNA [33,45], an occurrence that can be interpreted as a reduction of the persistence length b. Consequently, the equivalent spring constant $k_e$, Eq. (2), increases, and so does the force. In summary, this model supports the following interpretation of the increase in force observed in the experiments. In the initial stage, the $Ag^+$ ions make the dsDNA more flexible (smaller persistence length), and consequently, more force is required to keep it stretched by OTs.

As further confirmation of the dependence of the force on the persistence length, we performed Brownian simulations of a coarse-grained dsDNA model represented as a worm-like chain (WLC). Simulations on a reduced system with N = 500 beads, where the two terminal beads are constrained by a harmonic potential to mimic the two OTs, whose distance is 95% of the polymer contour length (i.e., a condition similar to the experimental case), show that the force decreases for more flexible polymers, see Fig 3b.

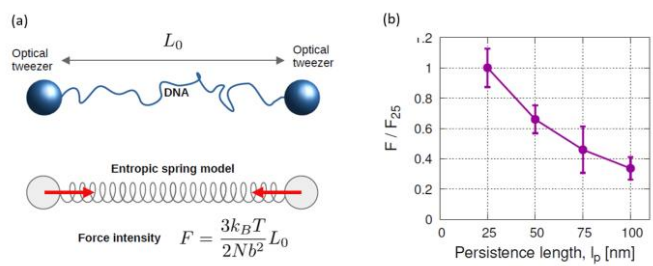

**Figure 3**. Entropic spring model. a) The experimental system can be modelled as an entropic spring, represented by the DNA anchored to two beads fixed in position by the optical tweezer. The ideal chain polymer model predicts that the force on the beads is proportional to their distance and increases at smaller Kuhn length $b$, i.e. for more flexible polymers. b) Brownian simulation of a worm-like-chain polymer model. Also in this case, the force is larger for smaller persistence lengths. The simulation uses a shorter polymer (contour length 2.5 μm) with respect to the λ-DNA in the experiments, but the same stretching (ratio between contour length and $L_0$). Forces are normalized with respect to the value $F_{z,25}$ corresponding to the $l_p$ = 25 nm case.

## Conclusions

The formation dynamics of DNA-templated silver nanowires were studied. Using OT experiments, it was found that the addition of 0.05 mM $Ag^+$ resulted in a 5.5-7.5 pN increase in force within 2 minutes, indicating that $Ag^+$ compresses the DNA structure. In contrast, the addition of hydroquinone caused the force to decrease by 4-5 pN within 2 minutes, thus allowing a stretching of the previously compressed DNA-$Ag^+$ structure. The morphologies of DNA-$Ag^+$ complex and DNA-Ag complex were examined using AFM, revealing the coverage of metal ions and metal atoms on the DNA during the formation of silver nanowires and metal atoms on DNA and conformational changes of DNA during the formation of silver wires. Furthermore, the dependency of the force on the persistence length was analyzed through Brownian simulations of a coarse-grained dsDNA model. In the future, optical tweezers could serve as a powerful tool for studying the mechanical properties and binding mechanics of other DNA-templated nanowires. This includes nanowires based on different metals, binding methods with metals, or special DNA structures, such as DNA origami. These results suggest that controlling the formation of nanowires may be feasible, opening up a wide range of potential applications in fields ranging from biology to nanoscience.

## Materials and methods

For the nanomechanical measurements we used λ-DNA biotinylated at its 5'-3' ends (Lumicks, Amsterdam, the Netherlands). For AFM imaging, we used λ-DNA from New England Biolabs (NEB). In all experiments, DNA was diluted to a final concentration of 0.04 μg/ml. 4.34 μm diameter streptavidin-coated polystyrene microbead (Spherotech, Chicago, USA) was diluted to a final concentration of .0.05%W/V. Figure S4a illustrates a tether formation principle between two polystyrene beads. Sodium phosphate buffer (PBS, pH=7.2), ammonium hydroxide ($NH_3OH$), (3-aminopropyl) triethoxysilane (APTES), $AgNO_3$, hydroquinone were purchased from Sigma and used without further purification. Purified water was obtained from a Milli-Q system (Miaozhiyi, Nanjing, China). PBS buffer was used as experimental buffer solution for DNA and microbeads. $NH_3OH$(Ph=10.5) was used as buffer solution for $AgNO_3$ and hydroquinone.

**Optical tweezer experiment**

Experiments were performed by using a dual-trap OTs instrument coupled with a five-laminar flow channel microfluidic system (Lumicks, M-Trap, Amsterdam, the Netherlands). The assembly protocol starts with two Streptavidin-coated polystyrene beads that are captured in the "beads channel" with the optical traps, and a biotinylated λ-DNA is then attached between two microbeads in the "tether channel", once the assembly is formed, it is important to ensure that only a single DNA was present between two beads. Because multiple DNAs influence force change which will mislead measurements. Figure S4(c-d) demonstrate force-extension curves for multiple tethers and single tether without any metallization, respectively. During DNA metallization, we forced on the dynamics of metallization for a single molecule, so the existence of single tether is confirmed by comparing with the worm-like chain(WLC) model. To obtain the force-time curve of kinetics of a single DNA metallization, the applied force (The distance between the beads was adjusted and kept constant) was adjusted before proceeding to metallize DNA. Subsequently, the molecule was brought to the "$AgNO_3$ Channel". When the change in force is stable, the whole assembly is finally moved to the "hydroquinone Channel" until the change in force is stable again. To alleviate hydrodynamic perturbations on DNA, flow was halted during the force measurement.

**Atomic force microscopy imaging**

Morphologies of the DNA-$Ag^+$ and the DNA-AgNPs were collected using an Atomic Force Microscope (AFM, Horiba, Japan), operated in tapping mode. Typical scan speeds were around 0.5 μm/s. Scanning resolution was 256 pixels/line for all images.
To enhance the adsorption of DNA, silicon substrates should be cleaned and treated by the APTES solution before depositing DNA. The silicon substrates were first treated by plasma cleaner (100W, 5min, $O_2$) to make the silicon extremely hydrophilic. Those silicon substrates were then modified in 100 μL of a 0.001%(v/v, ethanol) APTES solution for 10min, washed several times with 20 mL of ethanol to remove all APTES residues and to get uniform distribution of reactive amino groups, which enhance the interaction with DNA. Afterword, they were dried with nitrogen gas and were placed in a drying oven for 45minutes at 120 ℃. Figure S5 shows Procedure of the AFM sample preparation.

Before sample deposition, the λ-DNA was loaded with silver ions using a 0.05 mM $AgNO_3$ basic aqueous solution for 10, 20 minutes. The $Ag^+$-DNA complex was reduced with 0.002 mM hydroquinone solution for 10 min to form small metallic silver aggregates that were bound to the DNA skeleton. The environment should be dark during sample preparation. For each sample deposition, approximately 6 μl of the final solution was then dropped onto the silicon surface for 3 min, The surface was then cleaned 3 times with 20 μL of distilled water, followed by the removal of any excess solution using filter paper and dried with a gentle nitrogen gas flow before being subjected to AFM.

## Coarse-Grained Simulations

DNA is modelled as an almost inextensible polymer chain of N beads of mass m and position $\mathbf{r}_i$. We used a worm-like chain (WLC) model and the potential energy describing the system is

$$U(\mathbf{r}_1,\ldots,\mathbf{r}_N) = \frac{K_h}{2a^2}\sum_{i=1}^{N-1}(|\mathbf{r}_{i+1}-\mathbf{r}_i|-b)^2 - \frac{k_B T l_p}{a^3}\sum_{i=2}^{N-1}(\mathbf{r}_{i+1}-\mathbf{r}_i)(\mathbf{r}_i-\mathbf{r}_{i-1}) \quad (3)$$

where the first term is the sum of a stiff harmonic potentials such that each bond length can only perform negligible fluctuations around its equilibrium value, $|\mathbf{r}_{i+1}-\mathbf{r}_i| \cong \boldsymbol{\alpha}$ (almost inextensible condition). The second term represents the penalty that the chain has to pay upon bending, it enforces a chain in the bulk to maintain, on average, a persistence length, $l_p$. Note that, in general, the bending potential is equivalently expressed as $k_B T l_p a^{-1} \sum_{i=2}^{N-1}(\mathbf{n}_{i+1}\cdot\mathbf{n}_i)$, where $\mathbf{n}_{i+1}$ is the unit vector from bead $i$ and $i+1$.[46] However, since our polymer is almost inextensible, we use Eq. (3) that is computationally less expensive. The model is solved with an in house code already employed in ref.[47]

## Acknowledgements

We are grateful for continued financial support from National Natural Science Foundation of China (No. 22202167), National Key Research and Development Project of China (No. 2023YFF0613603), Provincial Science and Technology Plan Project: Micro and Nano Preparation and Photoelectronic Detection (No. 03014/226063). G.P.A and D.G. thank the European Union Program HORIZON-Pathfinder-Open: 3D-BRICKS, grant Agreement 101099125. D.G. and R.K. thanks the European Union under the Horizon 2020 Program, FETOpen: DNA-FAIRYLIGHTS, Grant Agreement 964995.

## Notes and references

1  K. C. Kao, W. Hwang and S. Choi, *Physics Today*, 1983, **36**, 90.

2  N. Fardian-Melamed, L. Katrivas, G. Eidelshtein, D. Rotem, A. Kotlyar and D. Porath, *Nano Lett.*, 2020, **20**, 4505–4511.

3  J. T. Petty, J. Zheng, N. V. Hud and R. M. Dickson, *J. Am. Chem. Soc.*, 2004, **126**, 5207–5212.

4  Z. V. Reveguk, E. V. Khoroshilov, Andrey. V. Sharkov, V. A. Pomogaev, A. A. Buglak and A. I. Kononov, *J. Phys. Chem. B*, 2024, **128**, 4377–4384.

5  Z. Reveguk, R. Lysenko, R. Ramazanov and A. Kononov, *Phys. Chem. Chem. Phys.*, 2018, **20**, 28205–28210.

6  A. Gonzàlez-Rosell, C. Cerretani, P. Mastracco, T. Vosch and S. M. Copp, *Nanoscale Adv*, 2021, **3**, 1230–1260.

7  R. R. Ramazanov, T. S. Sych, Z. V. Reveguk, D. A. Maksimov, A. A. Vdovichev and A. I. Kononov, *J. Phys. Chem. Lett.*, 2016, **7**, 3560–3566.

8  J. Lee, P. Lee, H. Lee, D. Lee, S. S. Lee and S. H. Ko, *Nanoscale*, 2012, **4**, 6408–6414.

9  P. Peng, W. Guo, Y. Zhu, L. Liu, G. Zou and Y. N. Zhou, *Nano-Micro Lett.*, 2017, **9**, 26.

10  P. Zhang, Q. Sui, Z. Liu, C. Hu, C. Li, X. Guo, J. Wang and G. Cai, *Chemical Engineering Journal*, 2024, **498**, 155277.

11  M. Kang, H. Lee, S. Hong and J. Choi, *Extreme Mechanics Letters*, 2023, **63**, 102059.

12  X. Jin, S. Kannappan, N. D. Hapsari, Y. Jin, K. K. Kim, J. H. Lee and K. Jo, *Small Structures*, 2023, **4**, 2200361.

13  Z. Chen, Y. Lin, C. Zhao, J. Ren and X. Qu, *Chem. Commun.*, 2012, **48**, 11428–11430.

14  K. Zhao, Q. Chang, X. Chen, B. Zhang and J. Liu, *Materials Science and Engineering: C*, 2009, **29**, 1191–1195.

15  H.-C. Hsu, M.-C. Ho, K.-H. Wang, Y.-F. Hsu and C.-W. Chang, *New J. Chem.*, 2015, **39**, 2140–2145.

16  L. Wu, J. Wang, J. Ren and X. Qu, *Advanced Functional Materials*, 2014, **24**, 2727–2733.

17  J. Lu, L. Yang, A. Xie and Y. Shen, *Biophysical Chemistry*, 2009, **145**, 91–97.

18  Q. Zhang, D. Jiang, C. Xu, Y. Ge, X. Liu, Q. Wei, L. Huang, X. Ren, C. Wang and Y. Wang, *Sensors and Actuators B: Chemical*, 2020, **320**, 128325.

19  R. Mahato, Sk. Masiul Islam and S. Singh, *Materials Today Communications*, 2024, **40**, 110056.

20  Q. Che, Q. Zhao, M. Hu, R. Qin, G. Shan and J. Yang, *ACS Appl. Nano Mater.*, 2021, **4**, 12726–12736.

21  E. Braun, Y. Eichen, U. Sivan and G. Ben-Yoseph, *Nature*, 1998, **391**, 775–778.

22  Z. Chen, L. Zhou, A. Zhao, Z. Zhang, Z. Wang, Y. Lin, J. Ren and X. Qu, *Biosensors and Bioelectronics*, 2014, **58**, 214–218.

23  S. Rao, S. Raj, S. Balint, C. B. Fons, S. Campoy, M. Llagostera and D. Petrov, *Applied Physics Letters*, 2010, **96**, 213701.

24  A. Woo, K. Lim, B. H. Cho, H. S. Jung and M.-Y. Lee, *Analytical Science Advances*, 2021, **2**, 397–407.

25  M. Dadmehr, M. Hosseini, S. Hosseinkhani, M. Reza Ganjali and R. Sheikhnejad, *Biosensors and Bioelectronics*, 2015, **73**, 108–113.

26  D. E. DiRico, P. B. Keller and K. A. Hartman, *Nucleic Acids Res*, 1985, **13**, 251–260.

27  R. M. Izatt, J. J. Christensen and J. H. Rytting, *Chem Rev*, 1971, **71**, 439–481.

28  J. Kondo, Y. Tada, T. Dairaku, H. Saneyoshi, I. Okamoto, Y. Tanaka and A. Ono, *Angewandte Chemie International Edition*, 2015, **54**, 13323–13326.

29  T. Dairaku, K. Furuita, H. Sato, J. Šebera, K. Nakashima, J. Kondo, D. Yamanaka, Y. Kondo, I. Okamoto, A. Ono, V. Sychrovský, C. Kojima and Y. Tanaka, *Chemistry – A European Journal*, 2016, **22**, 13028–13031.

30  X. Xing, Y. Feng, Z. Yu, K. Hidaka, F. Liu, A. Ono, H. Sugiyama and M. Endo, *Chemistry – A European Journal*, 2019, **25**, 1446–1450.

31  J. Kondo, Y. Tada, T. Dairaku, Y. Hattori, H. Saneyoshi, A. Ono and Y. Tanaka, *Nature Chem*, 2017, **9**, 956–960.

32  U. Javornik, A. Pérez-Romero, C. López-Chamorro, R. M. Smith, J. A. Dobado, O. Palacios, M. K. Bera, M. Nyman, J. Plavec and M. A. Galindo, *Nat Commun*, 2024, **15**, 7763.

33  A. A. Zinchenko, D. Baigl, N. Chen, O. Pyshkina, K. Endo, V. G. Sergeyev and K. Yoshikawa, *Biomacromolecules*, 2008, **9**, 1981–1987.

34  Z.-Y. Yang, W.-Y. Jiang and S.-Y. Ran, *Phys. Chem. Chem. Phys.*, 2023, **25**, 23197–23206.

35  D. A. Maksimov, V. A. Pomogaev and A. I. Kononov, *Chemical Physics Letters*, 2017, **673**, 11–18.

36  M. Fortino, T. Marino, N. Russo and E. Sicilia, *Phys Chem Chem Phys*, 2016, **18**, 8428–8436.

37  A. Ono, S. Cao, H. Togashi, M. Tashiro, T. Fujimoto, T. Machinami, S. Oda, Y. Miyake, I. Okamoto and Y. Tanaka, *Chem Commun (Camb)*, 2008, 4825–4827.

38  H. Mei, I. Röhl and F. Seela, *J. Org. Chem.*, 2013, **78**, 9457–9463.

39  N. Lefringhausen, C. Erbacher, M. Elinkmann, U. Karst and J. Müller, *Bioconjugate Chem.*, 2024, **35**, 99–106.

40  I. L. Volkov, A. Smirnova, A. A. Makarova, Z. V. Reveguk, R. R. Ramazanov, D. Y. Usachov, V. K. Adamchuk and A. I. Kononov, *J Phys Chem B*, 2017, **121**, 2400–2406.

41  J. Zhang, Y. Ma, S. Stachura and H. He, *Langmuir*, 2005, **21**, 4180–4184.

42  S. H. Park, M. W. Prior, T. H. LaBean and G. Finkelstein, *Applied Physics Letters*, 2006, **89**, 033901.

43  M. D. Wang, H. Yin, R. Landick, J. Gelles and S. M. Block, *Biophysical Journal*, 1997, **72**, 1335–1346.

44  M. Doi and H. See, in *Introduction to Polymer Physics*, eds. M. Doi and H. See, Oxford University Press, 1995.

45  W.-Y. Jiang and S.-Y. Ran, *The Journal of Chemical Physics*, 2018, **148**, 205102.

46  P. S. Doyle and P. T. Underhill, in *Handbook of Materials Modeling: Methods*, ed. S. Yip, Springer Netherlands, Dordrecht, 2005, pp. 2619–2630.

47  S. Zeng, M. Chinappi, F. Cecconi, T. Odijk and Z. Zhang, *Nanoscale*, 2022, **14**, 12038–12047.